\documentclass[preprint,nofootinbib]{revtex4}
\usepackage{graphicx}
\usepackage{amsmath,amssymb}

\newcommand{\be}{\begin{equation}}
\newcommand{\ee}{\end{equation}}
\newcommand{\bea}{\begin{eqnarray}}
\newcommand{\eea}{\end{eqnarray}}

\newcommand{\feyn}[1]{#1\!\!\!\!\slash\  }

\newcommand\bef{\begin{figure}}
\newcommand\eef[1]{\label{fg:#1}\end{figure}}
\newcommand\beq{\begin{equation}}
\newcommand\eeq[1]{\label{#1}\end{equation}}
\newcommand\beqa{\begin{eqnarray}}
\newcommand\eeqa[1]{\label{#1}\end{eqnarray}}
\newcommand\bet{\begin{table}}
\newcommand\eet[1]{\label{tb:#1}\end{table}}

\newcommand\fgn[1]{Figure \ref{fg:#1}}
\newcommand\eqn[1]{Eq.\ (\ref{#1})}
\newcommand\scn[1]{Section \ref{sec:#1}}

\begin{document}

\date{\today}

\title{
Transition in the spectral gap of the massless overlap Dirac operator coupled to abelian fields in three dimensions
}
\author{Rajamani\ \surname{Narayanan}}
\email{rajamani.narayanan@fiu.edu}
\affiliation{Department of Physics, Florida International University, Miami, FL 33199}

\begin{abstract}
The low lying spectrum of the massless overlap Dirac operator coupled to abelian fields in three dimensions with three different measures are shown to exhibit two phases: a strong coupling gapped phase and a weak coupling gapless phase. The vanishing of the gap from the strong coupling side with a Maxwell and a conformal measure is governed by a Gaussian exponent. Contrary to this result,  the vanishing of the gap from the strong coupling side with a compact Thirring measure is not consistent with a Gaussian exponent. 
The low lying spectrum with a non-compact Thirring measure does not exhibit a simple non-monotonic behavior as a function of the lattice size on the weak coupling side.
Our combined analysis suggests exploring the possibility of a strongly coupled continuum theory starting from a compact lattice Thirring model where a compact U(1) gauge field with a single link action is coupled to even number of flavors of massless overlap Dirac fermions.
\end{abstract}

\maketitle 

\section{Introduction}\label{sec:intro}

Strongly coupled theory of massless fermions interacting with an abelian gauge field in three dimensions (two space and one Euclidean time) has been studied over several decades both analytically and numerically~\cite{Semenoff:1984dq,Pisarski:1984dj,Appelquist:1985vf, Appelquist:1986fd, Appelquist:1986qw,
Appelquist:1988sr,Hands:1989mv,Hands:2002dv,Hands:2004bh,Hands:2006dh,Armour:2011zx,Karthik:2015sgq,Karthik:2016ppr,Karthik:2017hol,Karthik:2018tnh,Karthik:2019mrr,Karthik:2020shl} and has attracted recent attention due to its relevance in condensed matter physics and three dimensional duality~\cite{Miransky:2015ava,CastroNeto:2009zz,Franz:2002qy,Herbut:2002yq,Seiberg:2016gmd,Xu:2015lxa,Karch:2016sxi,Hsin:2016blu,Wang:2017txt}.
Another possibility for a theory of strongly interacting fermions in three dimensions is the Thirring model which is shown to be renormalizable in the limit of large number of flavors~\cite{Parisi:1975im,Hikami:1976at,Yang:1990ki,Gomes:1990ed,Hands:1994kb}  and a quest to find a strongly interacting continuum theory away from the limit of large number of flavors has been underway for several decades~\cite{DelDebbio:1995zc,Hands:1996px,DelDebbio:1997dv,DelDebbio:1999he,Hands:1999id,Christofi:2007ez,Christofi:2007ye,Gies:2010st,Janssen:2012pq,Schmidt:2015fps,Wellegehausen:2017goy,Hands:2017hhk,Hands:2018vrd,Hands:2018kvr,Lenz:2019qwu,Hands:2020itv}.
Unlike QED$_3$, the Thirring model after one converts the four fermi interaction into a bilinear with an auxiliary vector field is not gauge invariant. The interaction between fermions and the vector field is formally gauge invariant before regularization and one could perform an integral over all gauge transformations of the action for the vector field, namely $\left(\int d^3 x \ A_k^2(x)\right)$,  and enforce gauge invariance of the action by stating that only gauge invariant observables will be computed. Such a condition is usually not imposed when one converts the continuum action to a lattice regulated action for numerical analysis. At the outset, all recent numerical analysis to date~\cite{Schmidt:2015fps,Wellegehausen:2017goy,Hands:2017hhk,Hands:2018vrd,Hands:2018kvr,Lenz:2019qwu,Hands:2020itv} assume that fermions couple to a non-compact vector field with two different approaches treating the non-compact vector field on a site~\cite{Schmidt:2015fps,Wellegehausen:2017goy,Lenz:2019qwu} 
or a link~\cite{Hands:2017hhk,Hands:2018vrd,Hands:2018kvr,Hands:2020itv}. Recent numerical analysis of QED$_3$ has shown that theory remains scale invariant for all even number of flavors~\cite{Karthik:2015sgq,Karthik:2016ppr} as long as monopoles remain irrelevant. There is evidence for a critical number of flavors in the Thirring model~\cite{Hands:2018vrd,Hands:2020itv} if one uses domain wall fermions~\cite{Hands:2015qha,Hands:2015dyp} to regulate the fermions on the lattice and place the vector field on links. On the other hand, there is no evidence for symmetry breaking for any even number of flavors if one uses SLAC fermions to regulate the fermions on the lattice and place the vector field on sites~\cite{Wellegehausen:2017goy,Lenz:2019qwu}. Current conservation was used to show that all divergences in the $N$ flavor Thirring model 
up to order $\frac{1}{N}$ can be removed by standard renormalizations there by rendering it finite~\cite{Hands:1994kb}. This motivated us to see the effects of lattice regularization that preserves gauge invariance of the fermionic determinant.
 Toward that end we only focus on fermionic observables in a pure gauge measure, otherwise referred to as the quenched limit. The quenched limit of the Thirring model has been studied before~\cite{Hands:2018vrd} where evidence for symmetry breaking is found at sufficiently strong coupling but a possible transition to an unbroken phase at weak coupling was not fully explored since this limit is not physically interesting. Nevertheless, it is interesting from the view point of the studying the effects of different approach to lattice regularization. We will differ from~\cite{Hands:2018vrd} is one key aspect -- the lattice fermion operator will couple to the compact vector field thereby making it a gauge covariant interaction. We will use the overlap Dirac operator~\cite{Kikukawa:1997qh,Karthik:2016ppr} which is just a limit of the domain wall operator used in~\cite{Hands:2018vrd}. 

Consider the spectrum of the three dimensional Euclidean Dirac operator
\be
\feyn{D} = \sigma_k(\partial_k + iA_k);\qquad \sigma_1 = \begin{pmatrix} 0 & 1 \cr 1 & 0 \end{pmatrix};\quad
\sigma_2 = \begin{pmatrix} 0 & -i \cr i & 0 \end{pmatrix};\qquad \sigma_3 = \begin{pmatrix} 1 & 0 \cr 0 & -1 \end{pmatrix},
\ee
averaged over some measure, $P(A_k)$. Since the spectrum is invariant under the gauge transformation,
\be
A_k \to A_k + \partial_k \chi,
\ee
we can assume that the measure is also gauge invariant. We will also assume that the measure is invariant under parity, $A_k(x) \to - A_k(-x)$, and we will not concern ourselves with the parity anomaly. The following three choices of the measure are of relevance in the context of QED$_3$ and Thirring model:
\begin{itemize}
\item The Maxwell measure given by
\be
P_M(A) = \exp \left[ -\frac{\beta}{4} \int d^3 x F^2_{jk}\right];\qquad F_{jk} = \partial_j A_k - \partial_k A_j.\label{maxwell}
\ee
The spectrum is relevant in the quenched limit of QED$_3$ where the spectral density, $\rho(\lambda)$ , attains a non-zero value, $\rho(0)$, in the continuum limit, $\beta\to\infty$~\cite{Karthik:2017hol}.
\item The conformal measure given by
\be
P_C(A) = \exp \left[ -\frac{\beta}{4} \int d^3 x F_{jk}\frac{1}{\sqrt{-\partial^2}} F_{jk}\label{conformal}
\right].
\ee

The spectrum at various values of $\beta(N)$ can be matched with that of continuum parity invariant QED$_3$ with $2N$ flavors~\cite{Karthik:2020shl}.
\item The Thirring measure given by
\be
P_T(A) = \exp\left[ -\frac{\beta}{2} \int d^3x A_k^2\right] \quad {}^{{\rm gauge}\ {\rm averaging}}_{\qquad \Rightarrow\qquad } \quad \exp \left[ -\frac{\beta}{4} \int d^3 x F_{jk}\frac{1}{{-\partial^2}} F_{jk}\right].\label{thirring}
\ee
Analysis of the spectrum as a function of $\beta$ will shed some light into the existence of a strongly coupled theory in the continuum, albeit, in the quenched limit.
\end{itemize}

Our primary aim in this paper is to compare the spectrum of the overlap-Dirac operator on the lattice for the above three measures suitably discretized on the lattice.
On the one hand, the Maxwell measure is reasonably well understood due to the existence of a well defined continuum theory in the $\beta\to\infty$ limit. 
In addition, the behavior of the conformal theory with the conformal measure has been matched with the behavior of QED$_3$ with varying number of flavors~\cite{Karthik:2020shl}.
A comparison of the behavior with the Thirring measure with the other two measures can be used to explore the existence of a strongly interacting continuum theory at a finite value of $\beta$. 

Denoting the lattice link variables on a site of length $a$ by $\theta_k(x) =a A_k(x)$, the lattice overlap-Dirac operator will couple to the compact variable, $U_k(x) = e^{i\theta_k(x)}$.  It is natural to use a non-compact measure on the lattice for the Maxwell and the conformal case in order to suppress the presence of monopoles in the continuum theory.
But, we will consider the above Thirring measure and a variant since we are interested in eventually studying a field theory defined at a finite value of $\beta$.
Noting that fermions couple to $U_k(x)$, we rewrite the normalized measure for each $\theta_k(x)$ as~\cite{Drouffe:1983fv}
\bea
\sqrt{\frac{\beta}{2\pi}} \int_{-\infty}^\infty d\theta_k(x) e^{-\frac{\beta}{2} \theta^2_k(x)} &=&
\sqrt{\frac{\beta}{2\pi}}\int_{-\pi}^\pi d\theta_k(x) \left[ \sum_{n=-\infty}^\infty e^{-\frac{\beta}{2} \left( \theta_k(x)+2n\pi\right)^2} \right]\cr
&=&  \int_{-\pi}^\pi d\theta_k(x)\left[ \sum_{n=-\infty}^\infty \frac{e^{-\frac{n^2}{2\beta}}}{2\pi}  U_k^n(x)\right],\label{villain}
\eea
which is nothing but a {\sl Villain} type action~\footnote{The author would like to thank Simon Hands for bringing this to his attention.}. An alternative is to use a {\sl Wilson} type action for the link variables, namely,
\be
\frac{ e^{\frac{\beta}{2}\left (U_k(x) + U_k^*(x)\right)} }{2\pi I_0(\beta)}= \frac{e^{\beta\cos\theta_k(x)} }{2\pi I_0(\beta)} = \sum_{n=-\infty}^\infty \frac{I_n(\beta)}{2\pi I_0(\beta)} U_k^n(x).\label{thirringc}
\ee
The coefficients in the character expansion for the two choices reach the same limit as $\beta\to\infty$ but our interest is to look for a critical point at a finite value of $\beta$, possibly close to zero. Close to $\beta=0$, $I_n(\beta) = \frac{\beta^{|n|}}{2^{|n|} (|n|)!}$, and suppression of Wilson loop with a given size will be stronger in the {\sl Villain} type action compared to the {\sl Wilson} type action.

\section{Massless fermion spectrum on the lattice}

The massless overlap-Dirac operator, $D_o$, and the associated anti-Hermitian propagator, $A$, are~\cite{Karthik:2016ppr}
\bea
D_o = \frac{1+V}{2};\qquad V = X \frac{1}{\sqrt{X^\dagger X}};\qquad X = B+D;&&\cr
D = \frac{1}{2} \sum_{k=1}^3 \sigma_k(T_k-T^\dagger_k);\qquad B=\frac{1}{2}\sum_{k=1}^3 (2-T_k-T^\dagger_k) - m_w;&&\cr
(T_k \phi)(x) = e^{i\theta_k(x)}\phi(x+\hat k);&&\cr
A = \frac{1-V}{1+V}. &&
\eea
The Wilson mass parameter, $m_w$, can be taken anywhere in the range $(0,2)$ to realize a single massless fermion. Changing the value of $m_w$
will result in different values for the lattice spacing effects and we will set it to $m_w=1$ for all computations in this paper.
We will assume anti-periodic boundary conditions for fermions unless otherwise specified. Let the spectrum of $A^{-1}$ be defined by
\be
\frac{1+V}{1-V} \psi_j = i\Lambda_j \psi_j;\qquad j=\pm 1, \pm 2,\cdots, \qquad \cdots < \Lambda_{-2} < \Lambda_{-1} < 0 < \Lambda_1 < \Lambda_2 < \cdots.\label{ovspec}
\ee

\bef
\centering
\includegraphics[scale=0.58]{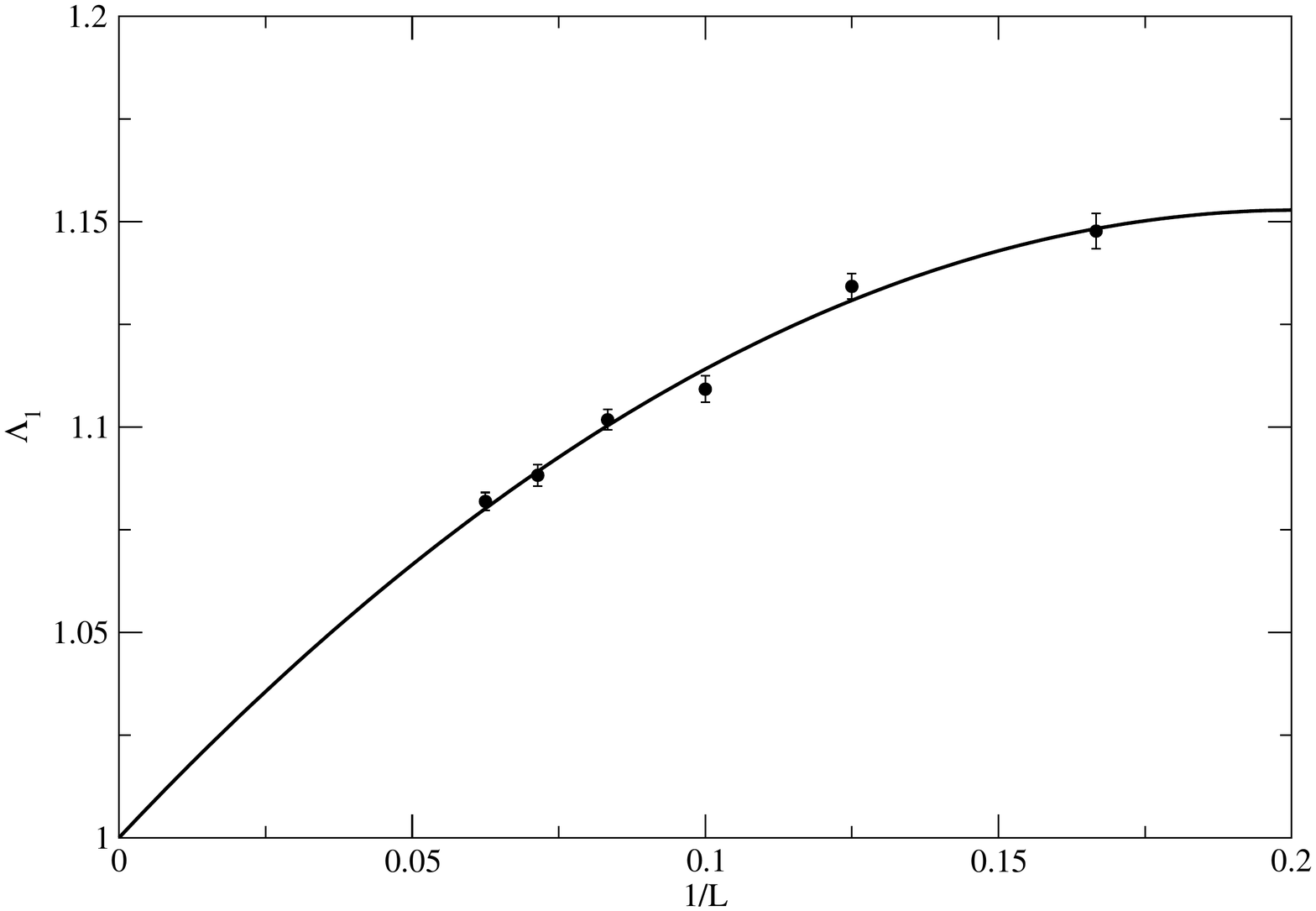}
\caption{A plot of $\langle \Lambda_1\rangle = \langle \Lambda_{-1}\rangle$ is shown as a function of $L$ in the strong coupling limit, $\beta=0$. The data is fitted to $1+\frac{a_0}{L} + \frac{a_1}{L^2}$
and shown to be consistent.
}
\eef{sc-gap}
In the weak coupling limit, $\beta = \infty$, all gauge actions discussed in \scn{intro} will result in gauge fields, $\theta_k(x)$, that are gauge equivalent to zero and the spectrum will be that of massless free fermions as expected. In the strong coupling limit, $\beta=0$, all gauge actions discussed in \scn{intro} will result in gauge fields, $\theta_k(x)$, that are uniformly and independently distributed in $[-\pi,\pi]$. The spectrum of the lattice Dirac operator will show a gap, namely, 
$
\lim_{L\to\infty} \langle \Lambda_{\pm 1} \rangle
$
will be consistent with unity and this corresponds to an eigenvalue of $V$ being $\pm i$. 
 It might be possible to obtain this result analytically using resolvents but it will be sufficient for our purposes to have shown it numerically by computing the spectrum on a few $L^3$ lattices and extrapolating to $L\to\infty$ as shown in \fgn{sc-gap}. 
To emphasize that the spectral gap is an effect at strong coupling, one could have repeated the calculation at $\beta=0$ but by coupling the fermions to a 
smeared link~\cite{Hasenfratz:2001hp,Hasenfratz:2007rf}.
Smearing takes a distribution of links that is uniform into one that favors $\theta_\mu(x)=0$. The effect on the gap will depend on the smearing parameters and the initial range of 
$\theta_k(x)$ which could be any real number instead of restricting it to $-[\pi,\pi]$.

\subsection{Spectral gap for a Maxwell measure}
\bef
\centering
\includegraphics[scale=0.29]{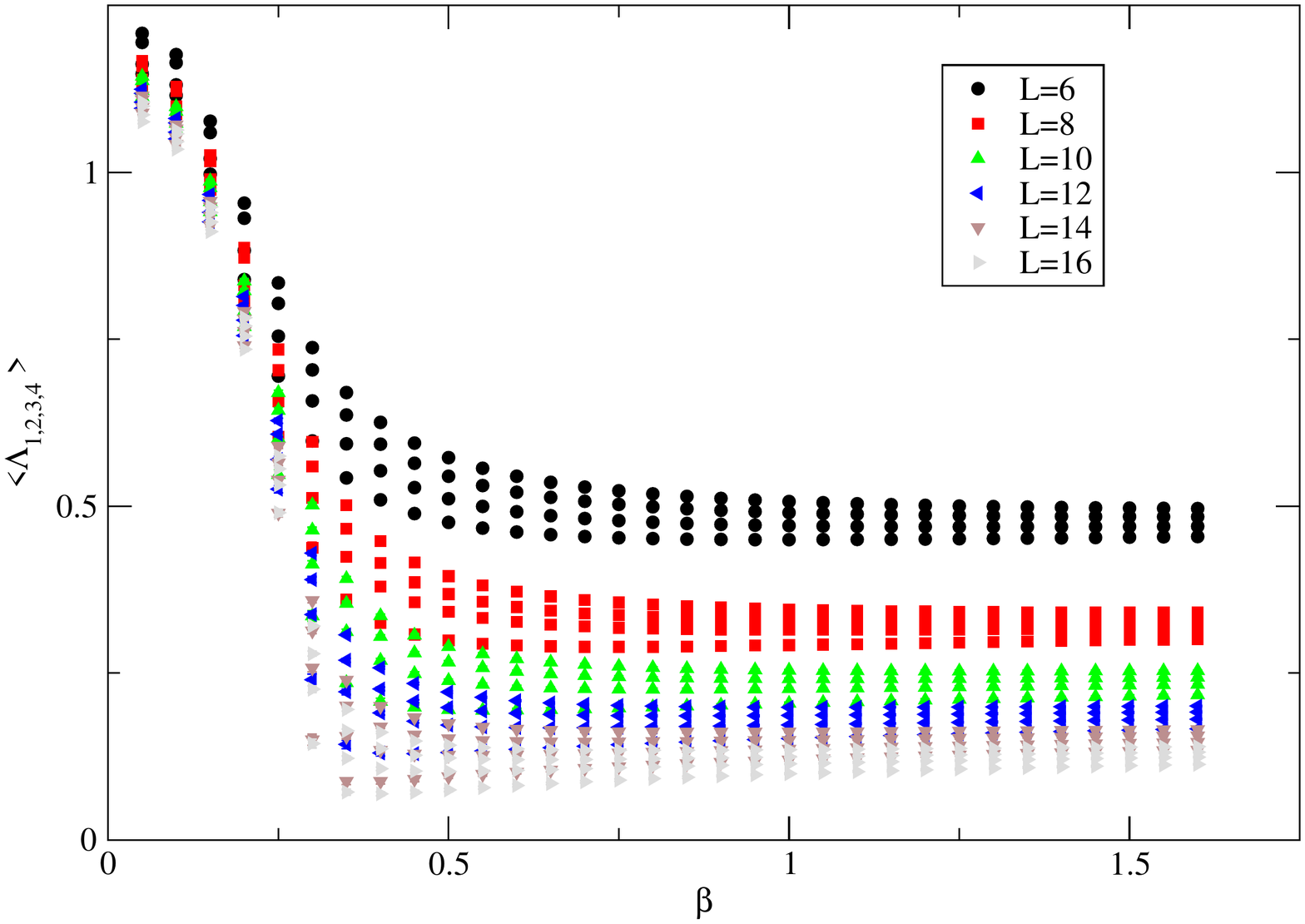}
\includegraphics[scale=0.29]{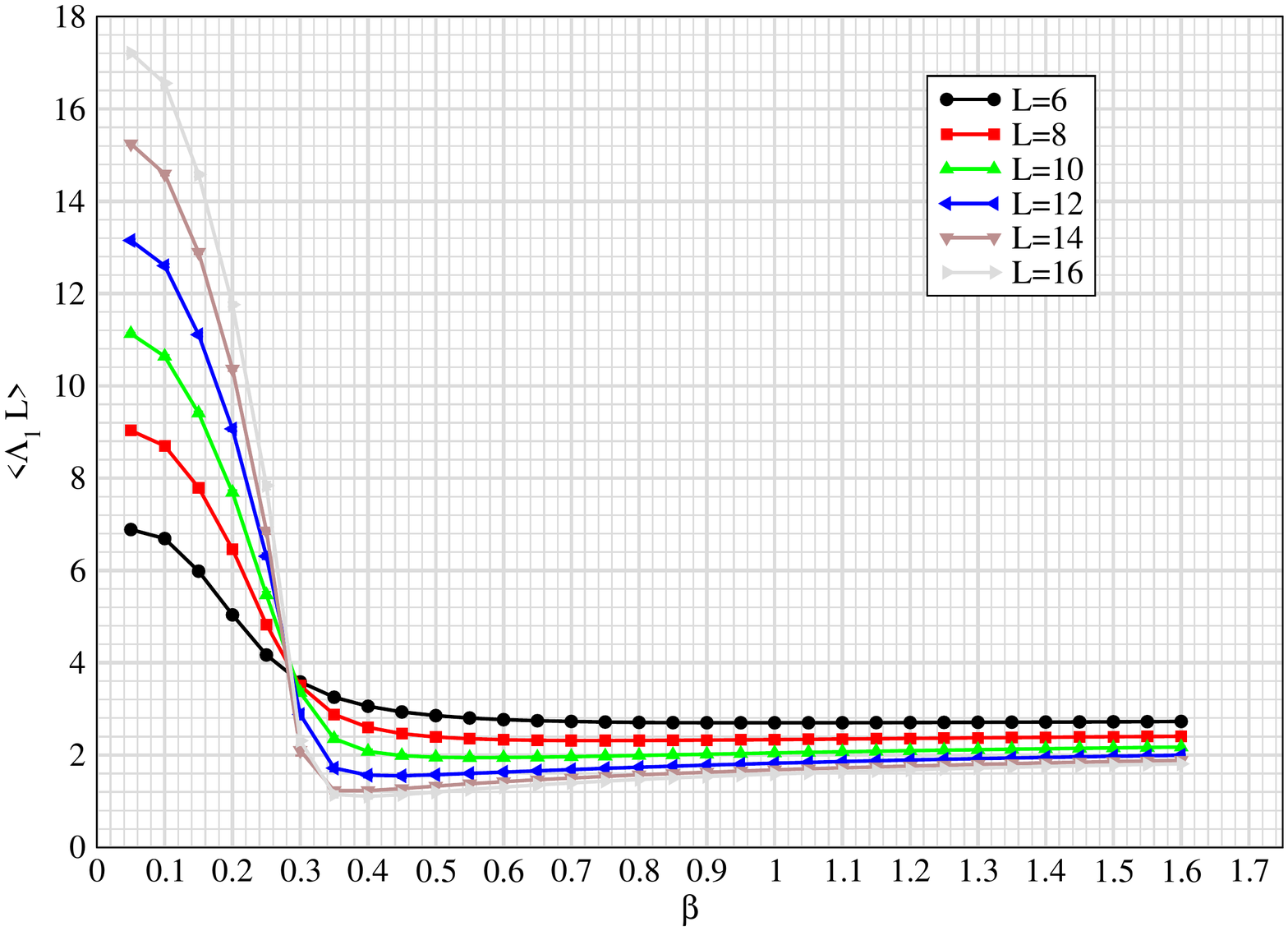}
\includegraphics[scale=0.58]{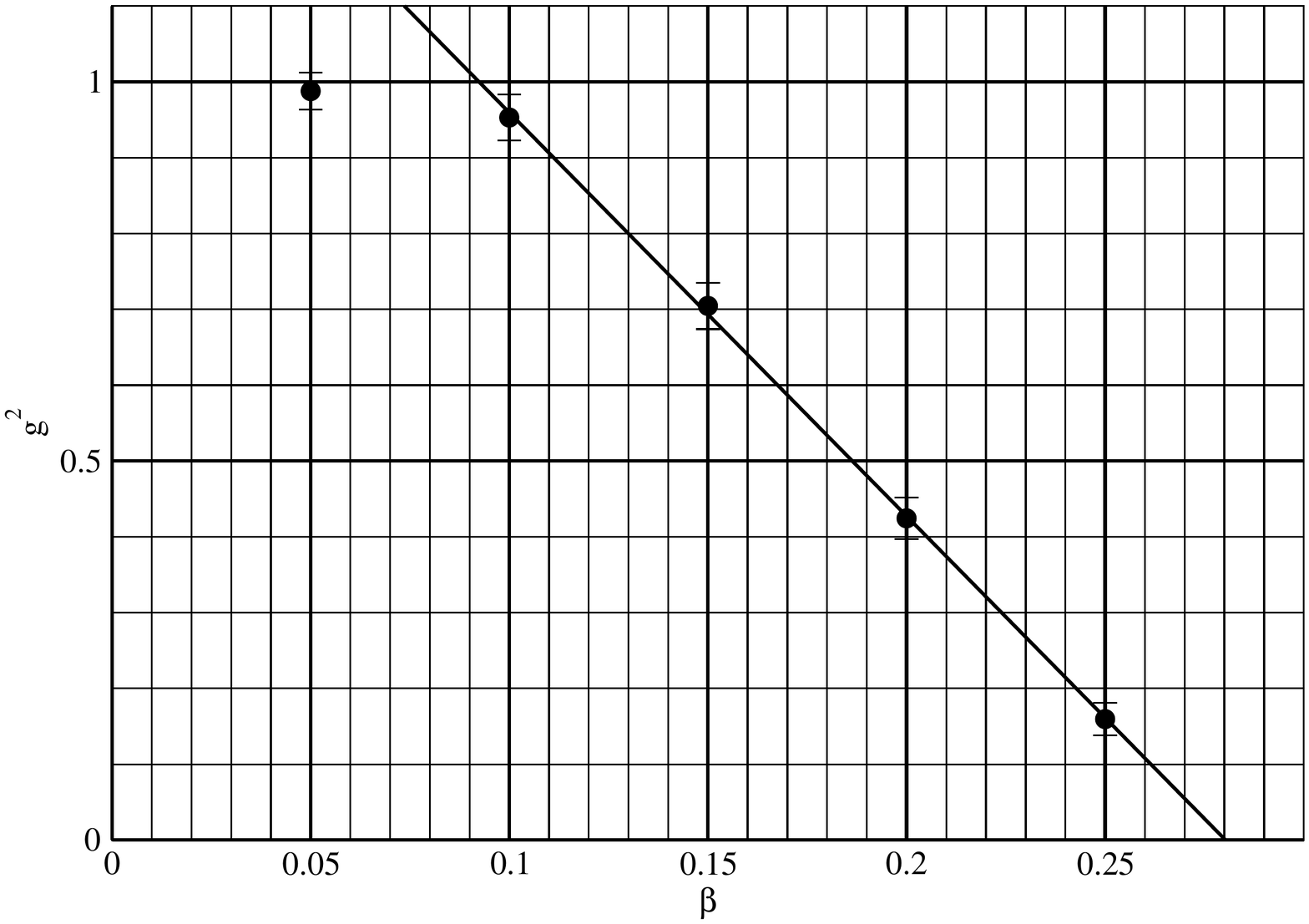}
\caption{The top left panel shows the low lying spectrum of the overlap Dirac operator with the Maxwell measure. The lowest scaled eigenvalue is shown on the top right panel and the presence of a critical coupling is indicated by the point where the curves at different values of $L$ cross each other. The bottom panel show the square of the gap extrapolated to $L=\infty$ for $\beta \le 0.25$ along with a simple linear regression.
}
\eef{max-gap}

The Maxwell measure given in \eqn{maxwell} translates to~\cite{Karthik:2020shl}
\be
\prod_n \exp \left [ -\frac{L^3 \beta f^2(n)}{2}\sum_{j=1}^2 \left|\tilde\theta_{j\perp}(n)  \right|^2\right]\label{maxmeas}
\ee
on a $L^3$ lattice where 
\be
\theta_k(x) = \sum_n \sum_{j=1}^2 \theta_{j\perp}(n) v_{j k}(n)\exp\left [ \frac{2\pi i n\cdot x}{L}\right]
\ee
and
\bea
f_j(n) = e^{\frac{2\pi i n_j}{L}} -1; && f(n)= \sqrt{\sum_{j=1}^3 |f_j(n)|^2};\cr
 \sum_{k=1}^3 v^*_{jk}(n) f_k(n) =0;\ j=1,2; &&
\sum_{m=1}^3 v^*_{jm}(n) v_{km}(n) = \delta_{jk}.
\eea
The zero mode ($n=0$) is not included in the measure. The low lying spectrum of the overlap Dirac operator, $\Lambda_i; i=1,2,3,4$ in \eqn{ovspec} was computed on $L=6,8,10,12,14,16$ for a range of coupling from $\beta=0.05$ to $\beta=1.6$ in steps of $\Delta\beta=0.05$ on $100$ independent configurations at each $L$ and $\beta$. The results are plotted in the top left panel of \fgn{max-gap}. The spectrum shows the discrete nature at finite volume and the eigenvalues themselves go down as $L$ increases. To show the presence of a transition from strong coupling to weak coupling, we plot the scaled lowest eigenvalue, $\Lambda_1 L$, in the top right panel of \fgn{max-gap}. There is clear evidence for a critical coupling, $\beta_c \approx 0.28$. The spectrum has a gap in the strong coupling side ($\beta < \beta_c$). The spectrum is gapless in the weak coupling side ($\beta > \beta_c$) since the scaled eigenvalue itself is going down as $L$ increases. The scaled eigenvalue on the weak coupling side falls faster away from $\beta\to\infty$ and this is consistent with the presence of a condensate that properly scales with $\beta$. A careful matching with random matrix theory should result in a condensate $\Sigma(\beta)$ for $\beta > \beta_c$.  We do not pursue this direction here since a value of the condensate has already been numerically computed~\cite{Karthik:2017hol}. In order to emphasize the presence of a gap on the strong coupling side, we extrapolated the lowest eigenvalue, $\Lambda_1$, as a function of $L$ to a gap at $L=\infty$ using a fit of the form $g+\frac{a_0}{L} + \frac{a_1}{L^2}$. The square of the gap, $g^2$, so obtained 
with errors obtained by single elimination jackknife has been plotted for $\beta\le 0.25$ in the bottom panel of \fgn{max-gap}. The square of the gap at $\beta=0.1,0.15,0.2,0.25$ fit a simple linear regression quite well with an estimate of the critical coupling consistent with $\beta_c \approx 0.28$. If this simple minded analysis survives further scrutiny, the transition from the strong coupling (gapped side) to weak coupling (gapless side) is a second order transition with Gaussian exponents.

\subsection{Spectral gap for a conformal measure}

\bef
\centering
\includegraphics[scale=0.29]{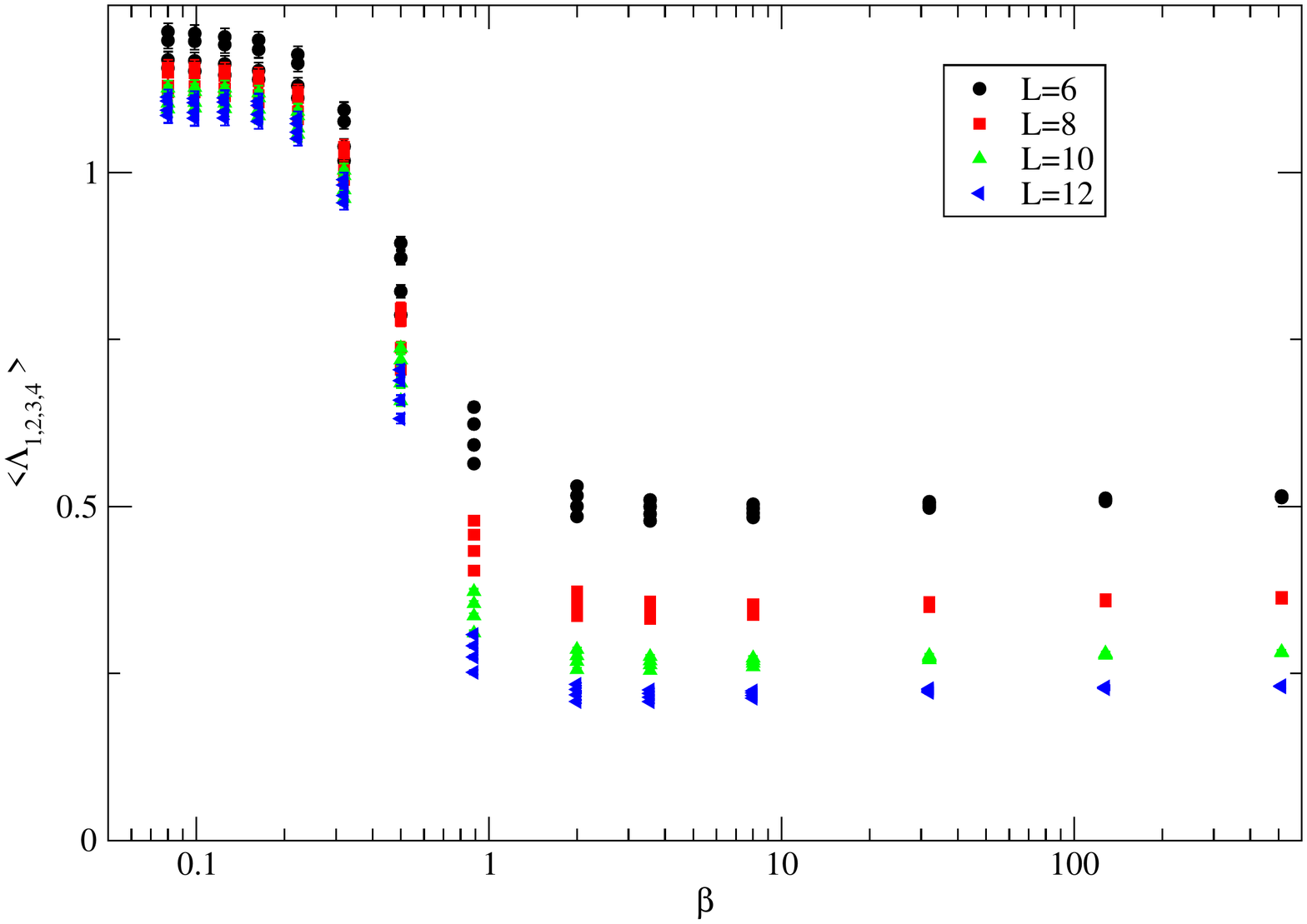}
\includegraphics[scale=0.29]{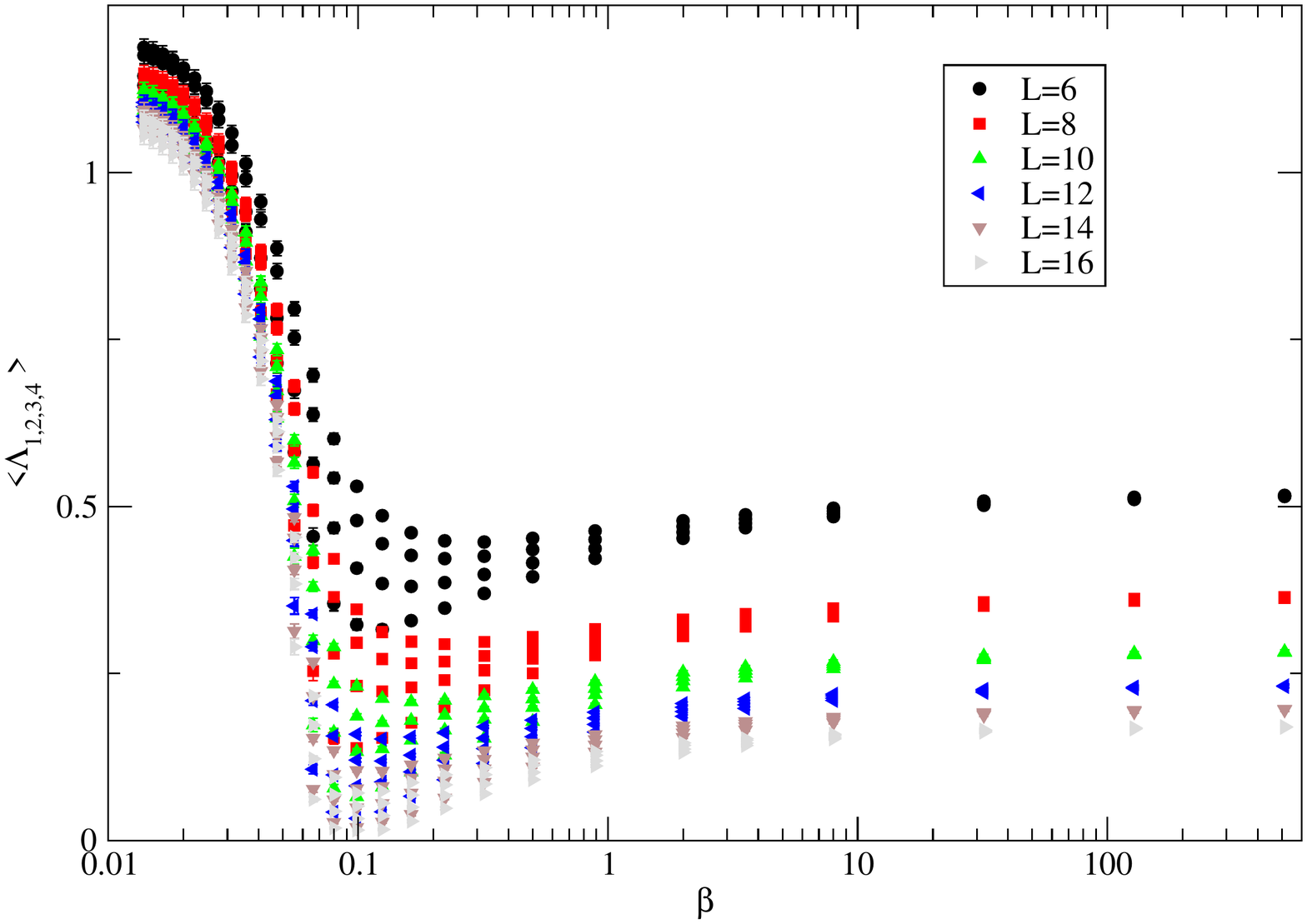}
\includegraphics[scale=0.29]{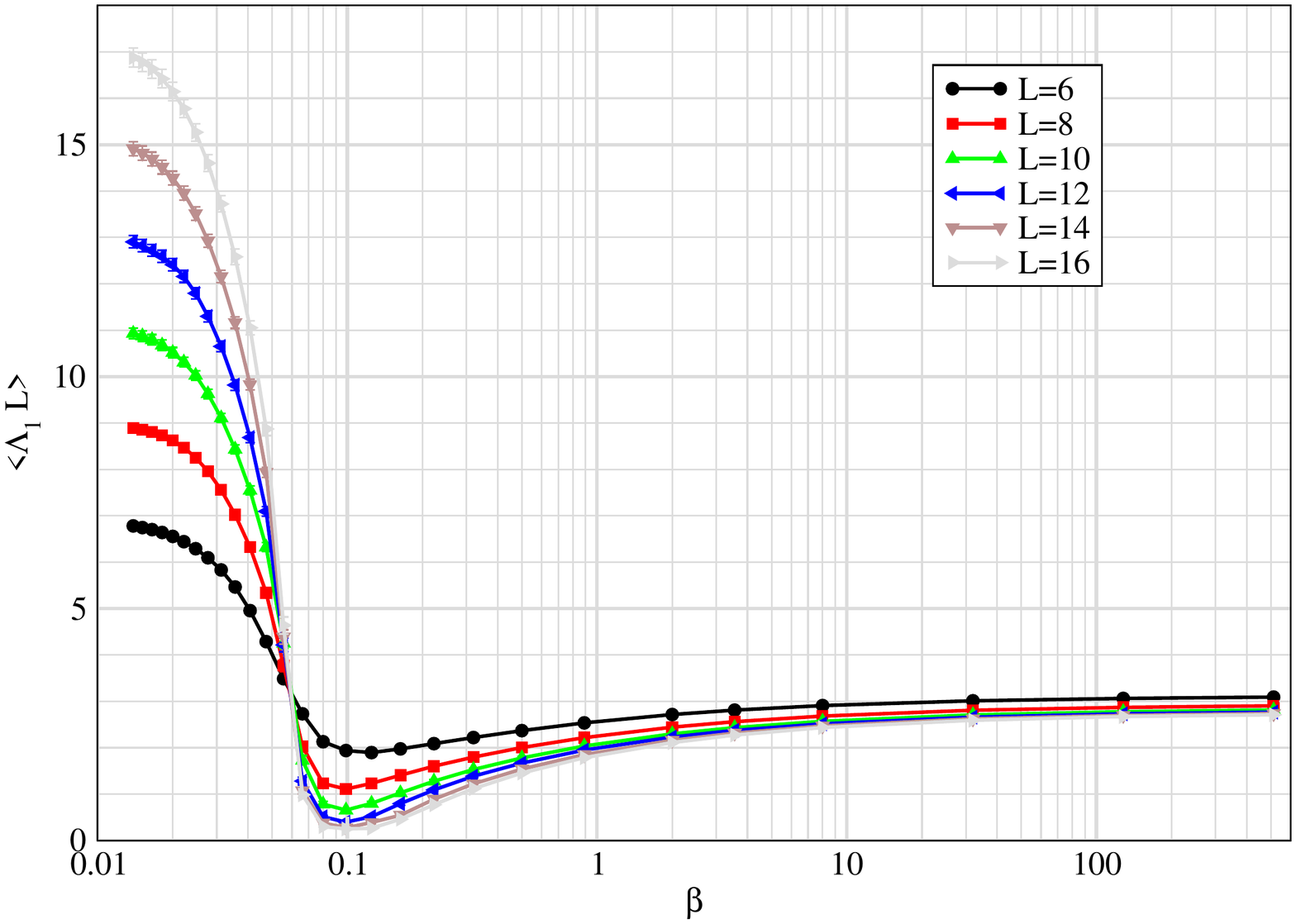}
\includegraphics[scale=0.29]{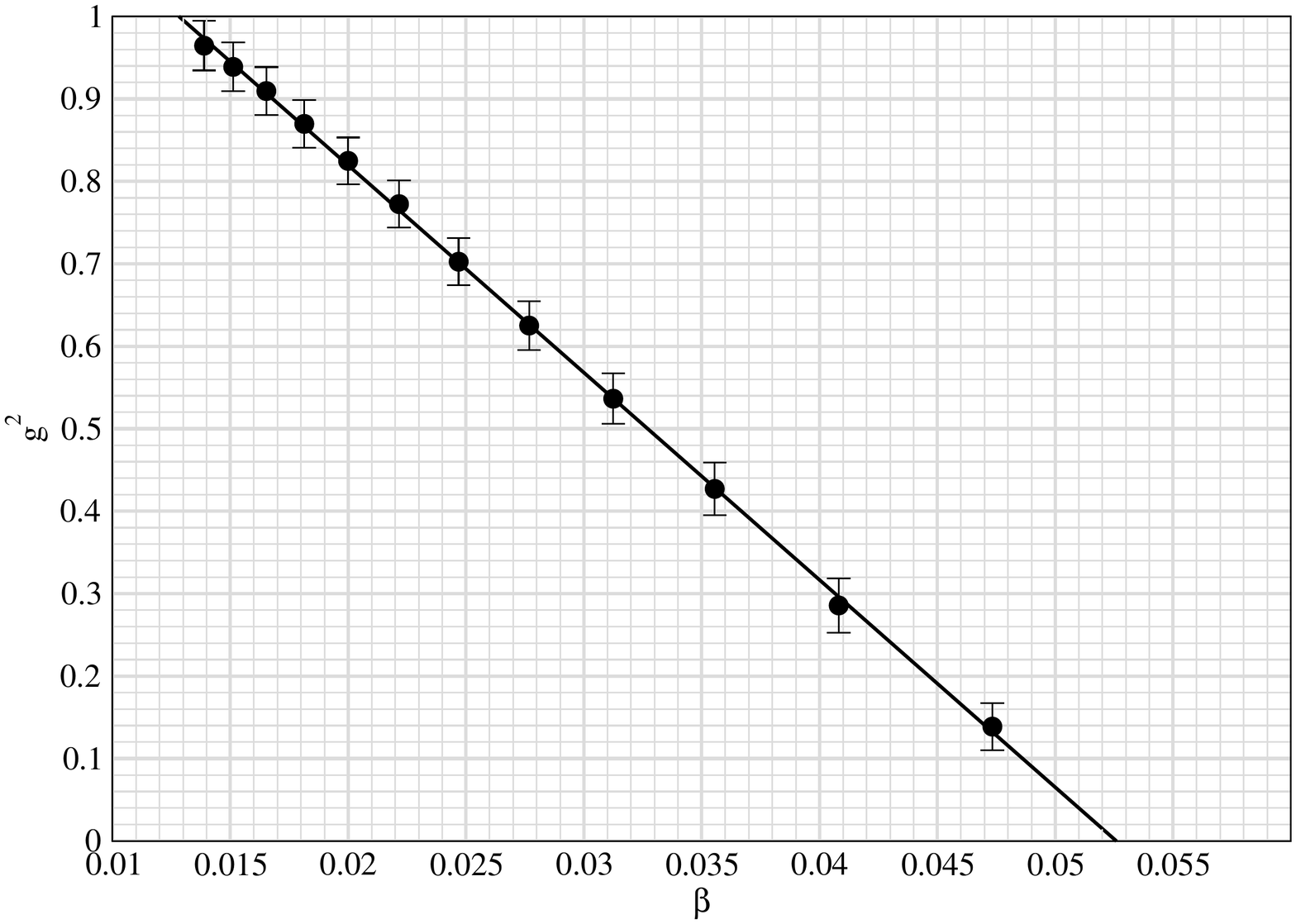}
\caption{The top left panel shows the low lying spectrum of the overlap Dirac operator coupled to unsmeared gauge fields generated with the conformal measure. The top right panel shows the low lying spectrum of the overlap Dirac operator coupled to smeared gauge fields generated with the conformal measure.The lowest scaled eigenvalue is shown on the bottom panel and the presence of a critical coupling is indicated by the point where the curves at different values of $L$ cross each other. The bottom right panel show the square of the gap extrapolated to $L=\infty$ for $\beta \le 0.25$ along with a simple linear regression. The overlap Dirac operator is coupled to the smeared gauge fields in both the bottom panels.
}
\eef{conf-gap}

The conformal measure given in \eqn{conformal} amounts to changing the weight of each mode in the Maxwell measure (c.f. \eqn{maxmeas}) to
\be
\prod_n \exp \left [ -\frac{L^3 \beta f(n)}{2}\sum_{j=1}^2 \left|\tilde\theta_{j\perp}(n)  \right|^2\right].\label{confmeas}
\ee
This measure is of relevance in computing fermionic observables and matching them to observables in QED$_3$ coupled to $N_f$ flavors (assumed to be even to preserve parity) of two component massless fermions~\cite{Karthik:2020shl}. The coupling $\beta$ becomes equal to $N_f$ when $N_f\to\infty$ but one can match $\beta(N_f)$ for the entire range of $N_f \ge 2$. To obtain the matching function, $\beta(N_f)$, the fermions were coupled to smeared gauge fields in~\cite{Karthik:2020shl} and we will study the effect of smearing on the spectral gap in this section.
 The low lying spectrum of the overlap Dirac operator coupled directly to the unsmeared gauge fields generated by \eqn{confmeas}, $\Lambda_i; i=1,2,3,4$ in \eqn{ovspec} were computed on $L=6,8,10,12$ for a range of coupling from $\beta=0.08$ to $\beta=512$ on $100$ independent configurations at each $L$ and $\beta$. The results are plotted in the top left panel of \fgn{conf-gap}. Like with the Maxwell measure, the spectrum shows the discrete nature at finite volume and the eigenvalues themselves go down as $L$ increases. 
A gap develops close to $\beta=1$ and this might hinder the matching, $\beta(N_f)$, over the entire range of $N_f$. If the gauge fields appearing in the overlap Dirac operator are smeared, the gap shifts to smaller $\beta$ and the results for $\Lambda_i; i=1,2,3,4$  computed on $L=6,8,10,12,14,16$ for a range of coupling from $\beta=0.0139$ to $\beta=512$ on $100$ independent configurations at each $L$ and $\beta$ are shown in the top right panel of \fgn{conf-gap}.
 To show the presence of a transition from strong coupling to weak coupling, we plot the scaled lowest eigenvalue, $\Lambda_1 L$, with smeared gauge fields in the bottom 
 left panel of \fgn{max-gap}. There is clear evidence for a critical coupling, $\beta_c \approx 0.05$. The spectrum has a gap in the strong coupling side ($\beta < \beta_c$). The spectrum is gapless in the weak coupling side ($\beta > \beta_c$) since the scaled eigenvalue itself is going down as $L$ increases. A continuum limit can be obtained at every value of $\beta> \beta_c$ with this measure and the scaling of the eigenvalue with $L$ at a fixed $\beta$ gives the anomalous dimension at that $\beta$. The anomalous dimension goes down as $\beta$ increases and this is consistent with the trend in the dependence of the eigenvalues with $L$ for $\beta > 0.1$. Only values of coupling, $\beta > 3.5$ were used to match with QED$_3$ in~\cite{Karthik:2020shl}. A careful analysis of the anomalous dimension for the entire range of $\beta > \beta_c$ might be interesting and might not even be a monotonic function of $\beta$. 
  In order to emphasize the presence of a gap on the strong coupling side, we extrapolated the lowest eigenvalue, $\Lambda_1$, as a function of $L$ to a gap at $L=\infty$ using a fit of the form $g+\frac{a_0}{L} + \frac{a_1}{L^2}$. The square of the gap, $g^2$, so obtained with errors obtained by single elimination jackknife has been plotted for $\beta\le 0.05$ in the bottom right panel of \fgn{conf-gap}. The square of the gap for $\beta \in [0.0139,0.0473] $ fit a simple linear regression quite well with an estimate of the critical coupling consistent with $\beta \approx 0.0525$. This shows the Maxwell measure and the conformal measure that describe the same physics also exhibit the same mean field behavior close to the critical lattice coupling where the gap closes.
  
\subsection{Spectral gap for a compact Thirring measure}

\bef
\centering
\includegraphics[scale=0.29]{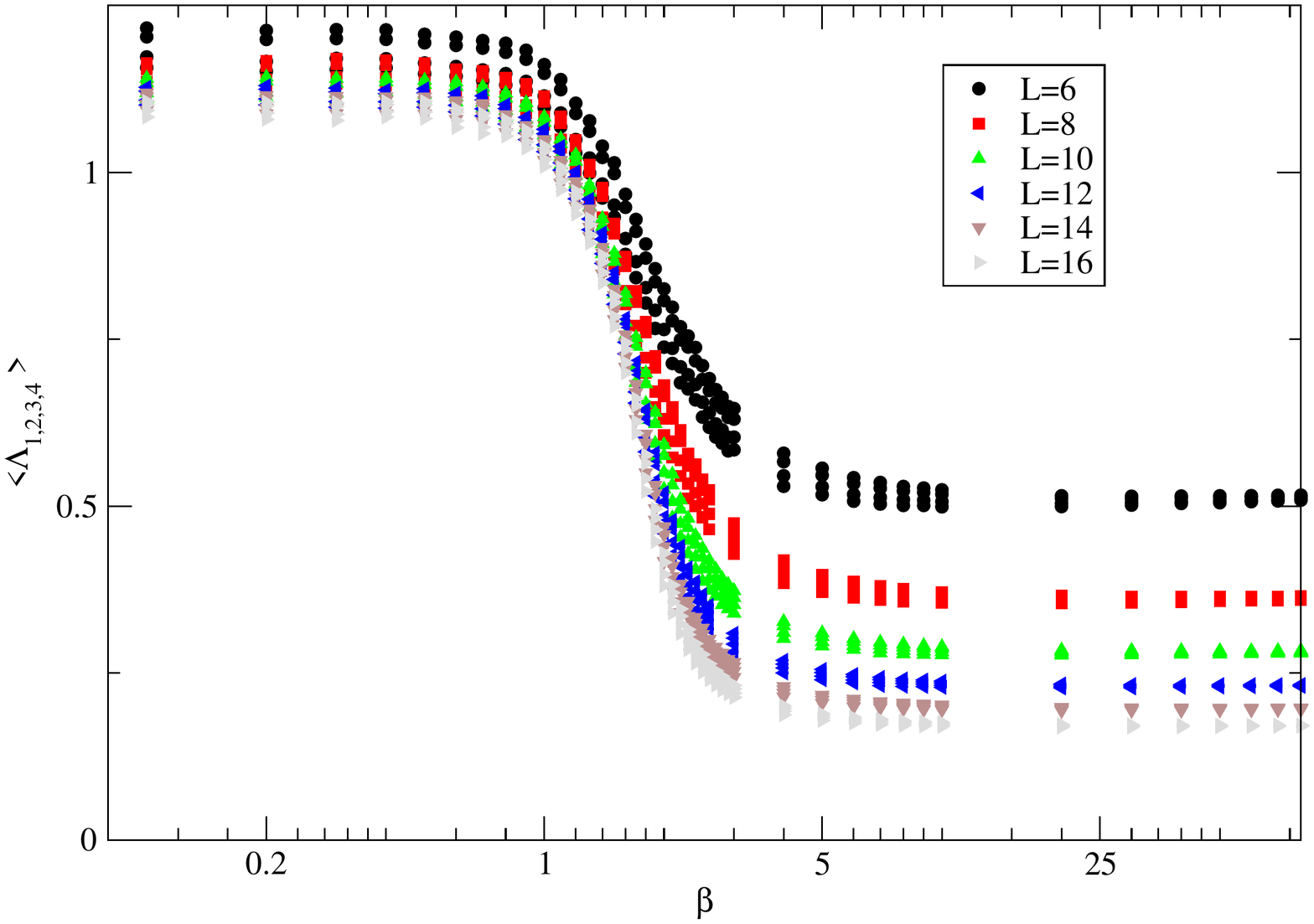}
\includegraphics[scale=0.29]{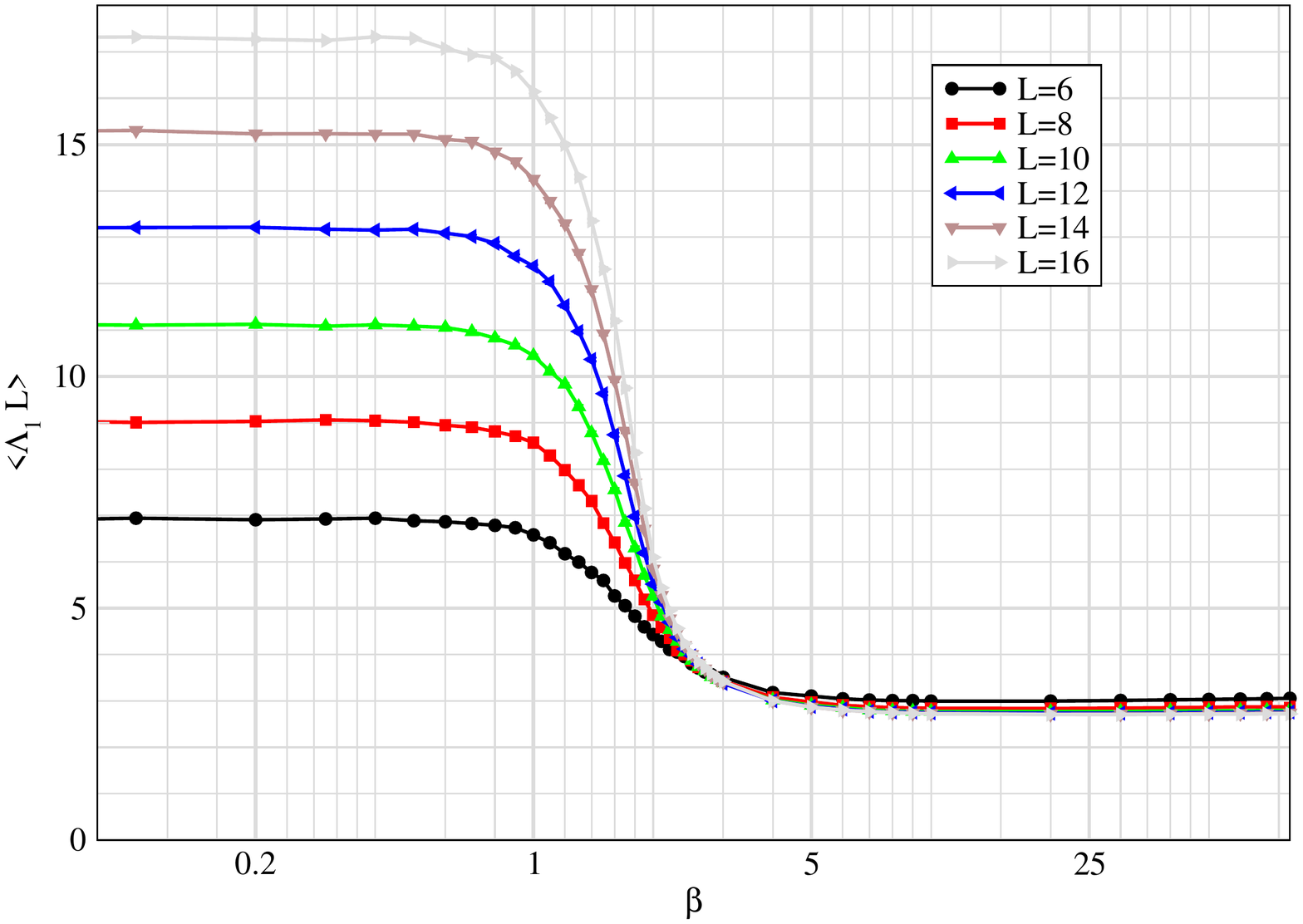}
\includegraphics[scale=0.29]{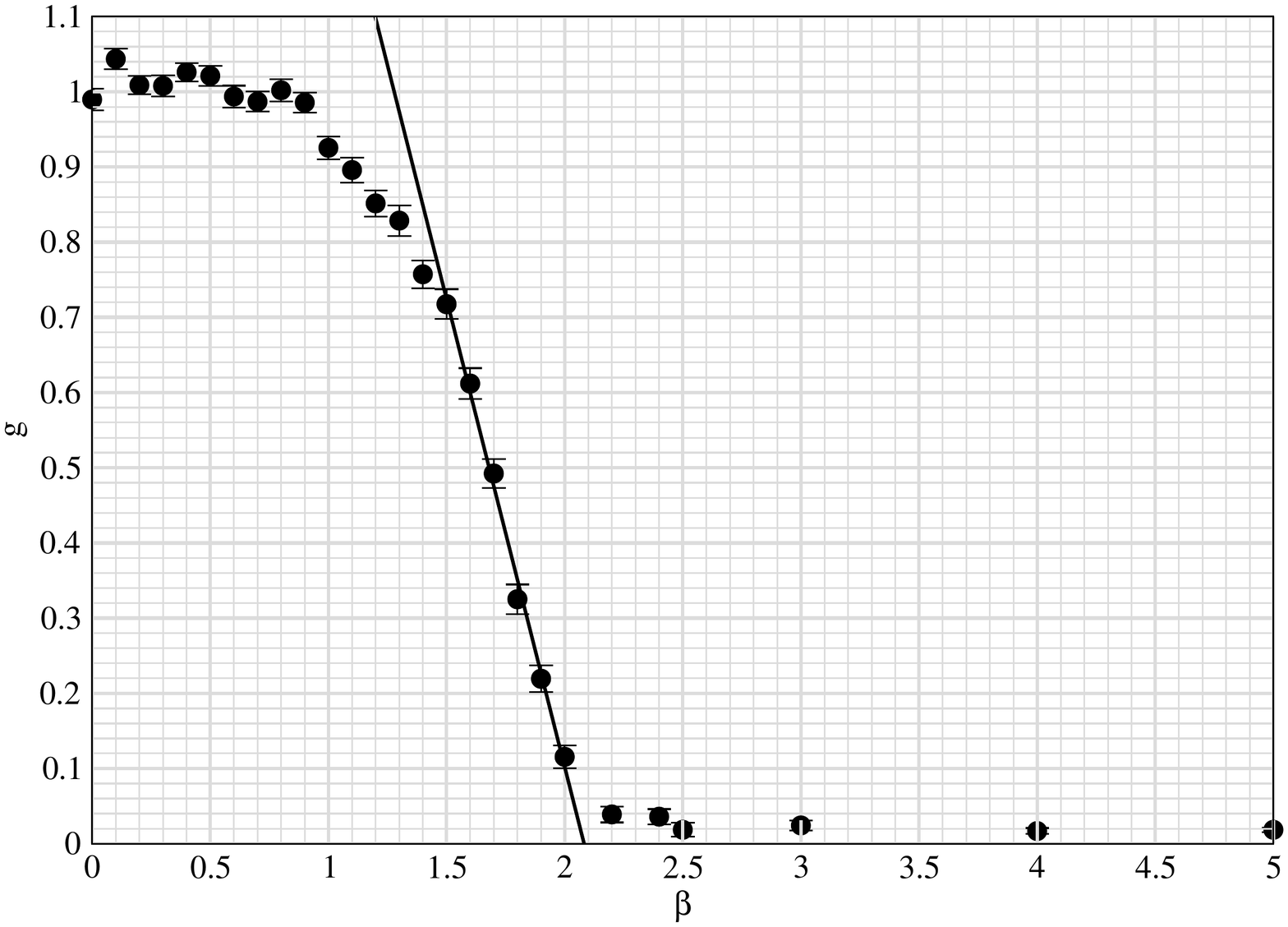}
\includegraphics[scale=0.29]{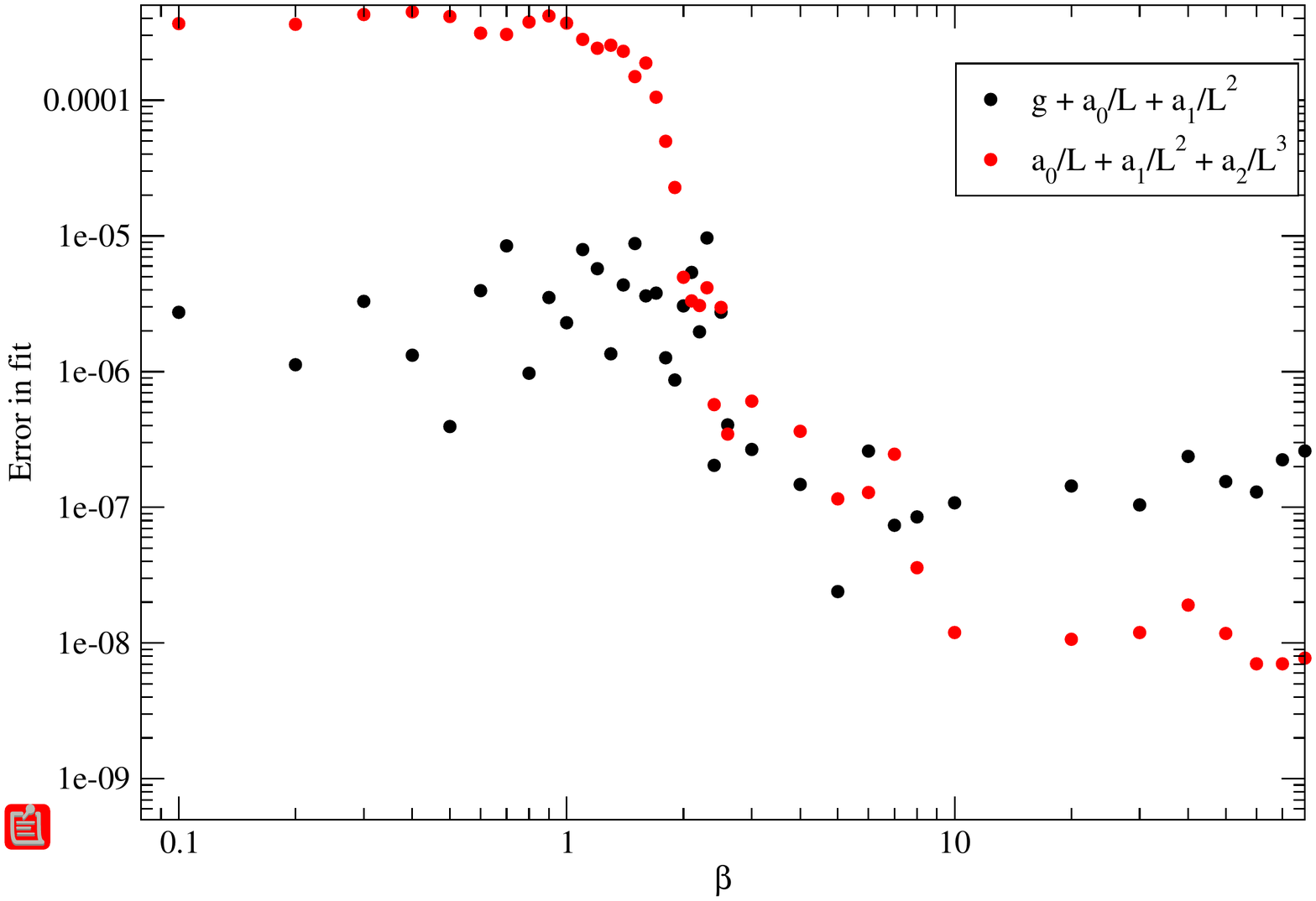}
\caption{The top left panel shows the low lying spectrum of the overlap Dirac operator coupled to gauge fields generated with the compact Thirring measure. The lowest scaled eigenvalue is shown on the top right panel and the presence of a critical coupling is indicated by the point where the curves at different values of $L$ cross each other. The bottom left panel show the gap extrapolated to $L=\infty$ for $\beta \le 5$ along with a simple linear regression over a region that covers $\beta\in [1.5,2]$. The
bottom right panel shows the error in the fit with $g\ne 0$ and $g=0$.}
\eef{thir-gap}

We now move on to the Thirring measure which is the main new point of this paper. Our aim is to compare the behavior of the low lying spectrum of the overlap Dirac operator
with the Thirring measure to the ones from the Maxwell and the conformal measure. To this end, we first focus on the compact Thirring measure given in \eqn{thirringc}.
  We generated $\theta_k(x)$ with a measure $\exp[\beta (\cos(\theta_k(x))-1)]$ independently for each $k$ and $x$ on a $L^3$ lattice.
The behavior of the four lowest eigenvalues as a function of $\beta$ are shown in the top left panel of \fgn{thir-gap} for six different values of $L$
and the qualitative similarity with the corresponding behaviors in \fgn{max-gap} with the Maxwell measure and \fgn{conf-gap} with the conformal measure is evident.
Like with the Maxwell and the conformal measures, we have plotted the scaled lowest eigenvalue, $\Lambda_1 L$ as a function of $\beta$ in the top-right panel of \fgn{thir-gap}.
There is evidence for a critical value around $\beta_c\approx 4$. Whereas the increase with increasing $L$ below $\beta_c$ is as pronounced as in the Maxwell and conformal measures, the behavior above $\beta_c$ seems to be essentially independent of $L$ and the value of $\Lambda_i L$ is close to $\pi$ indicating free field behavior. Contrary to this behavior, the free field behavior with the Maxwell measure has not set it even at $\beta=1.7$ and there is clear tendency for $\Lambda_1 L$ to go below $\pi$ as $L$ is increased. The free field behavior in $\Lambda_1 L$ does seem to set in with the conformal measure for large enough $\beta$ but the values stay well below $\pi$ for a wide range of $\beta$ along with a variation in the behavior as a function of $\beta$. Like in the Maxwell and conformal measures, we extrapolated $\Lambda_1$ to obtain a gap using the fit $g+\frac{a_0}{L} + \frac{a_1}{L^2}$. The result with errors obtained by single elimination jackknife is shown in the bottom left panel of \fgn{thir-gap}. There are qualitative differences when one compares the gap with the Thirring measure to the ones from the Maxwell and conformal measures. The square of the gap with the Maxwell and conformal measures fit a simple minded linear regression quite well for $g^2$ close to $1$ and close to $0$. On the other hand, we have plotted $g$ as a function of $\beta$ for the Thirring model and we see a region in the gap, namely, $[0.1,0.7]$, where a simple linear regression fits well. But this results in a critical value of $\beta_c\approx 2.1$ which is not consistent with the estimate from the behavior of $\Lambda_1 L$ as a function of $\beta$ for different values of $L$ in the top right panel of \fgn{thir-gap}.
To further understand this discrepancy, we also fitted $\Lambda_1$ to $\frac{a_0}{L} + \frac{a_1}{L^2} + \frac{a_2}{L^3}$ which assumes there is no gap.
The fit errors, namely the value of the least square function, from the two different fits are shown in the bottom right panel of \fgn{thir-gap}. It is clear that a non-zero gap is favored for $\beta < 2$.
It is also reasonably clear that a fit with no gap is favored for $\beta>8$. The region, $2 < \beta < 7$, is murky and it is difficult to conclude the presence or the absence of a gap with the data that is currently available. Combining this with a possible behavior of $g \sim (\beta-2.1)$ for $\beta\in [1.5,2]$, this model might have a lattice critical point which is characterized by non-Gaussian exponents.
 
\section{Why use a compact Thirring measure?}

\bef
\centering
\includegraphics[scale=0.29]{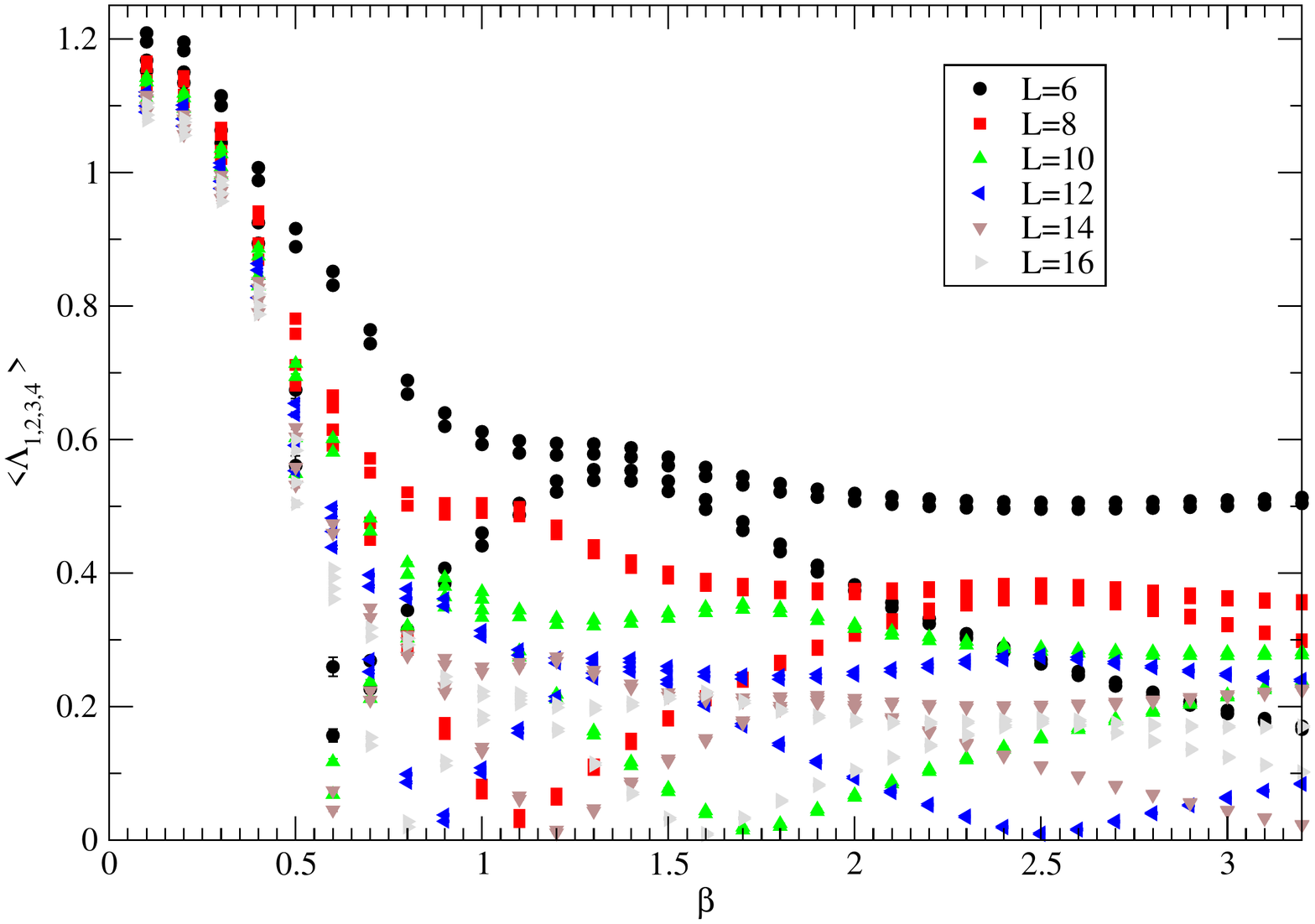}
\includegraphics[scale=0.29]{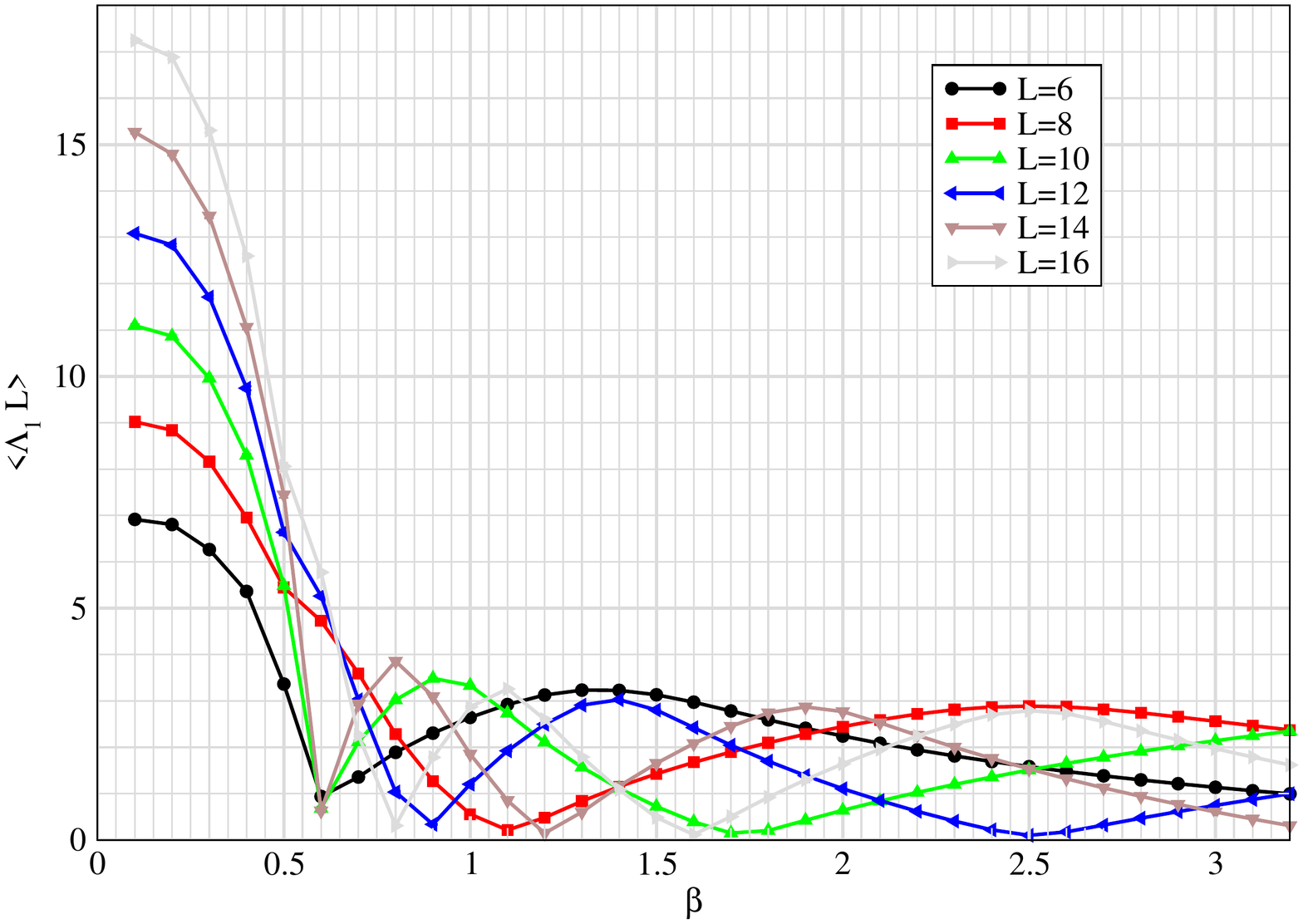}
\includegraphics[scale=0.29]{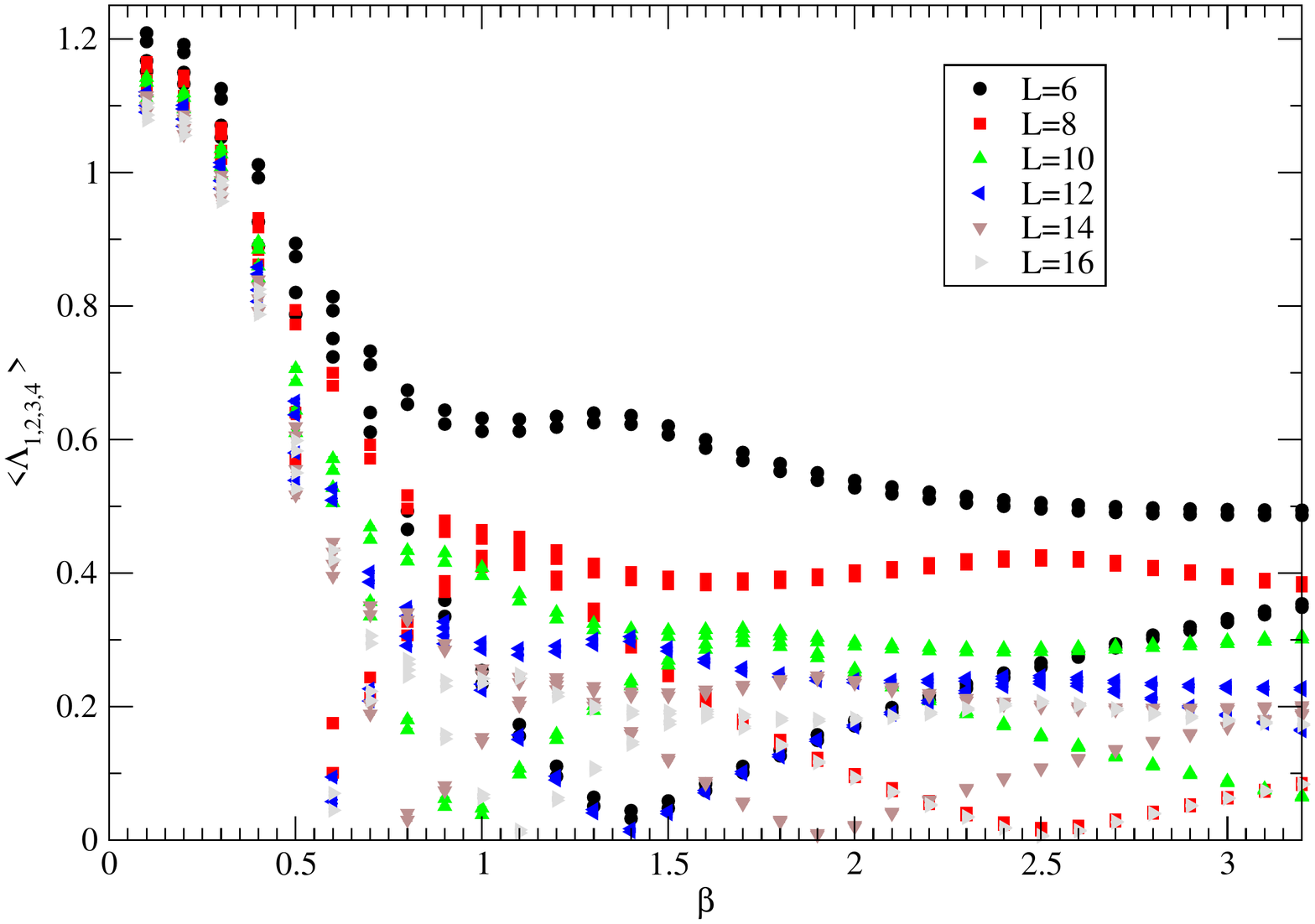}
\includegraphics[scale=0.29]{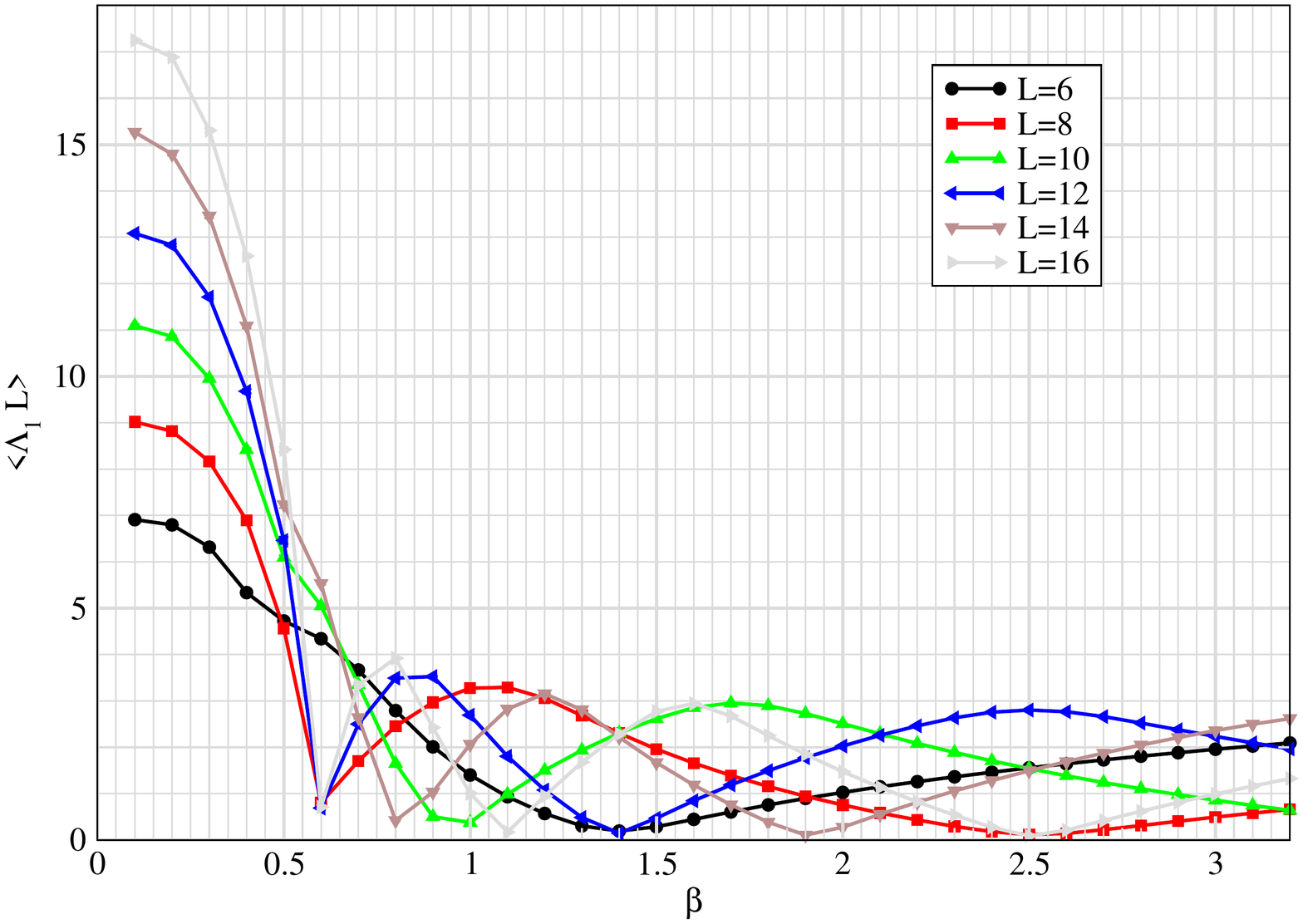}
\caption{The top and bottom left panels shows the low lying spectrum of the overlap Dirac operator coupled to gauge fields generated with the non-compact Thirring measure. The lowest scaled eigenvalue is shown on the top and bottom right panels. The fermions obey anti-periodic boundary conditions in the top panels whereas they
obey periodic boundary conditions in the bottom panels.
}
\eef{ncthir-gap}
The Thirring measure that arises from converting a four fermi interaction to a bilinear with an auxillary vector field is given by \eqn{thirring}. We generated $\theta_k(x)$
using a measure $\exp[-\frac{\beta}{2} \theta^2_k(x))]$ independently for each $k$ and $x$ on a $L^3$ lattice and studied the behavior of the low lying spectrum of the overlap Dirac operator. In this case, we computed the spectrum with both anti-periodic and periodic boundary conditions for fermions. The results with anti-periodic and periodic boundary conditions are shown in the top and bottom panels respectively. The monotonic behavior of the eigenvalues as a function of $L$ as a fixed $\beta$ seen in the strong coupling side is absent in the weak coupling side. The presence of a single critical coupling is not evident in the plots of the scaled lowest eigenvalue in the plots on the right panels. Since the fermion always sees the compact gauge field, the difference in the behavior between the non-compact gauge action (c.f. \eqn{villain})
and the compact gauge action (c.f. \eqn{thirringc}) is in suppression of larger Wilson loops in the non-compact action compared to the compact gauge action.
This difference in the actions plays a strong qualitative role in the behavior of the spectrum at intermediate coupling which is the region of interest in the Thirring model.

\section{Discussion}

We numerically studied the low lying spectrum of the overlap Dirac operator coupled to a compact Abelian gauge field. We investigated four different measures for the gauge field. All of them clearly show a lattice strong coupling phase where the spectrum is gapped: There are no eigenvalues in the range $i[-g,g]$; $g > 0$.
The natural interpretation of the gapped phase is the effect of a random gauge potential that persists up to a certain value of the coupling constant (see~\cite{Brezin:1994sq} for example). One can change the location of the critical coupling where the gap closes by coupling the overlap Dirac operator to a smeared gauge field; the transition from the gapped phase to a gapless phase is a lattice transition. The location of the lattice transition will also depend on the choice of the Wilson mass parameter, $m_w$, used in the overlap Dirac operator.
One expects to realize a continuum theory in the gapless phase. In order to see if a continuum theory can be realized in the three dimensional Thirring model, we compared the fermion spectrum with a compact Thirring measure to a Maxwell measure and a conformal measure.
The Maxwell measure is relevant for QED$_3$ and the conformal measure becomes relevant for QED$_3$ due to its scale invariant behavior.
All three measures show a clear separation of the two phases. There is reasonable evidence that the gap with the Maxwell and conformal measures go to zero as $\sqrt{\beta_c - \beta}$ from the strong coupling side as one approaches the critical coupling. This Gaussian like behavior is not seen with the Thirring measure where there is some indication that the gap behaves closer to $(\beta_c-\beta)$ from the strong coupling side. But the location of the transition is itself not well determined based on the data used in this paper and further work has to be done to explore the intermediate region of coupling and perform a careful finite volume analysis. We also studied the spectrum with the non-compact Thirring measure and the behavior of the low lying eigenvalues in the potentially gapless phase is non-monotonic in the size of the lattice. This effect is a consequence of suppressing Wilson loops of larger size and can be avoided by using the compact Thirring action. 

Our analysis suggests that it would be interesting to study the $2N$ flavor compact Thirring model with the lattice action including fermionic sources defined by
\be
S =\frac{\beta}{2} \sum_{k,x} \cos\theta_k(x) + \sum_{j=1}^{N} \left( \bar\psi_j D_o \psi_j  + \bar\psi_j D^\dagger_o \psi_j  \right) 
+ \sum_{j=1}^{2N} \bar\eta_j A \eta_j.\label{conthir}
\ee
A lattice model similar to this one with staggered fermions and an explicit integration over all gauge transformations of $\theta_k(x)$ was already proposed in~\cite{Kim:1996xza}.
In order to make a connection between the above action and the one for QED$_3$ with $2N$ flavors of massless fermions, consider the
generalized continuum action defined by
\be
S=\frac{2N}{4g^2} \int d^3 x (\partial_k A_j - \partial_j A_k)\left[\frac{1}{-\partial^2}\right]^p(\partial_k A_j - \partial_j A_k)  +
\int d^3 x \bar\psi_j \sigma_k (\partial_k+i A_k)\psi_j ,\label{maxtothir}
\ee
with the coupling constant, $g^2$, having a mass dimension of $(1-2p)$. The $p=0$ theory corresponding to QED$_3$ shows scale invariant infra-red behavior for all even number of flavors~\cite{Karthik:2015sgq,Karthik:2016ppr}. Furthermore the induced action for the gauge fields is well described by $p=\frac{1}{2}$ conformal theory with no dynamical fermions~\cite{Karthik:2020shl}. Theories with $p> \frac{1}{2}$ are not renormalizable by simple power counting but are shown to be well defined in the large $N$ limit~\cite{Parisi:1975im,Hikami:1976at,Yang:1990ki,Gomes:1990ed,Hands:1994kb}.
Strictly in the large $N$ limit, the beta function is exactly equal to
\be
\beta(\bar\alpha) = (2p-1)\bar\alpha(1-\bar\alpha),
\ee
where $\bar\alpha(p)$ is the dimensionless running coupling constant~\cite{Appelquist:1986fd}. This suggests that the infra-red and ultra-violet behavior are flipped when a theory with $p$ is compared to $(1-p)$. It would be interesting to see if one can realize a continuum theory with $p=1$ using \eqn{conthir} by first finding the location in the lattice coupling where the gap in the fermion spectrum closes and studying the scaling limit from the weak coupling side.
\acknowledgments
The author acknowledges partial support by the NSF under grant number
PHY-1913010 and would like to thank Simon Hands, Nikhil Karthik and Jude Worthy for several useful discussions.
\bibliography{../../mynotes/biblio}
\end{document}